\documentclass{phb-proc4-auth}

\usepackage{graphicx}
\usepackage{amssymb}

\def\beq{\begin{equation}}
\def\eeq{\end{equation}}
\def\eqref#1{Eq.~(\ref{#1}) }
\newcommand{\cal}{}

\def\LNSCO{La$_{1.6-x}$Nd$_{0.4}$Sr$_x$CuO$_{4}$}
\def\delt{ \delta t }
\newcommand{\eps}{\varepsilon}

\begin{document}
\begin{frontmatter}


\journal{SCES '04}

\title{Competition Between Charge-Density Waves and
 Superconductivity in 
Striped Systems}

%
%
%
%
%
%

 \author[TUG]{Enrico Arrigoni\corauthref{1}}
 \author[UI]{Eduardo Fradkin}
  \author[UCLA]{Steven A. Kivelson}

%
 
\address[TUG]{
Institute for  Theoretical Physics,
Graz University of Technology, Petersgasse 16, A-8010 Graz, Austria
}
\address[UI]{Department of Physics, 
University of Illinois at Urbana-Champaign, 1110 W. Green St., 
Urbana, IL 61801-3080}
\address[UCLA]{
Department of Physics, 
University of California at Los Angeles, 405 Hilgard Ave., 
Los Angeles, CA 90095
}

%
%
%
%


%
%
%
%

\corauth[1]{Corresponding Author: 
Graz university of Technology, Petersgasse 16, A-8010 Graz, Austria.
Phone +43-316-873-8180,
Fax:+43-316-873-8677,
Email:arrigoni@itp.tu-graz.ac.at
}


\begin{abstract}

Switching on interchain coupling in a system of one-dimensional
strongly interacting chains often leads to an ordered state.
Quite generally, there is a competition between an insulating
 charge-density-wave and a superconducting state. In the case of
 repulsive interactions, charge-density wave usually wins over
 superconductivity.

Here, we show that a suitable modulation in the form of a period $4$ 
bond-centered stripe can reverse this balance {\it even in the
  repulsive case} and produce a
superconducting state with relatively high temperature.

\end{abstract}

%
%

\begin{keyword}

Superconductivity \sep Quasi-one-dimensional systems \sep competing phases

\end{keyword}


\end{frontmatter}

%
%
%
%
%

There is 
 strong
 {\it theoretical} evidence
that quasi-one dimensional ladder systems with purely repulsive
electronic interaction, such as two-leg Hubbard or $t-J$ ladders, 
show a strong tendency toward the formation of a large spin
gap associated with
substantial superconducting (SC) ``d-wave-like'' pair-field
correlations~\cite{scal.nature,da.ri.science,ha.po.95.es,no.wh.94.ct},
i. e. the pairing susceptibility diverges as
\beq
\chi_{SC} \sim \Delta_s/T^{2-1/K} \;.
\eeq
Here, $\Delta_s$ is the spin gap, $T$ the temperature and $K$ the
charge Luttinger parameter of the ladder.
One could thus imagine to obtain a
 ``true'' superconductor with long-ranged
correlations  by coupling an infinite number
of ladders into a two-dimensional array.
However, 
besides SC, also charge-density-wave (CDW) 
susceptibility 
with wave vector $4K_F$
diverges with a power law in the temperature:
\beq
\chi_{CDW} \sim \Delta_s/T^{2-K} \;.
\eeq

\begin{figure}
     \centering
     \includegraphics[width=4cm]{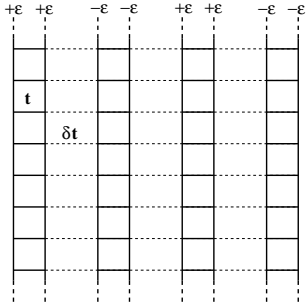}
     \caption{
\label{lad4}
Schematic representation of the coupled ladder system 
with alternating on-site energies
discussed in the text. The ladders can be described, e. g.,  by a single-band Hubbard
model or by a $t-J$ model. The ladders are then
coupled via a hopping $\delt\ll \Delta_s$, and an alternating on-site energy $\pm
\eps$ has been introduced in order to frustrate CDW pinning, as
discussed in the text.
}
 \end{figure}

Numerical calculations~\cite{ha.po.95,tr.ts.96,no.wh.96,no.bu.97} have shown that, 
except in a very small region in parameter space,
 $K$ is smaller than $1$
for  strong repulsive
 interactions, so that
 CDW susceptibility diverges stronger
 than 
 SC.
For this reason, 
upon switching on a weak
coupling between the ladders, one expects 
CDW correlations to
condense
 before the SC ones, so that
in the end one obtains 
a CDW
insulator instead of a superconductor.
This can be seen from the RG equations for the interchain CDW and
Josephson couplings, ${\cal V}$ and $\cal J$ respectively:
\beq
\label{rg}
{d \cal V/}{d \tau} = (2 - K) {\cal V} \quad\quad
{d {\cal J}/}{d \tau} = (2 - 1/K) {\cal J}  \;,
\eeq 
where $\tau $ is the RG parameter.
The ``starting'' values for the dimensionless couplings  at the energy scale of
the spin gap~\cite{dop} are given by ${\cal V}(\tau=0)= {\cal J}(\tau=0) 
\sim (\delt)^2/\Delta_s^2$, where $\delt$ is the microscopic interladder 
hopping parameter (see Fig.~\ref{lad4}).
\eqref{rg} shows that the CDW order parameter increases faster than
SC for $K<1$, and, thus, it is the first one to reach strong coupling.
The question is whether there is a way to reverse this result and 
have superconductivity
 ``win'' the competition against  CDW, by still keeping $K<1$.

The solution consists in 
introducing   a frustration in the CDW
periodicity between the ladders. 
To understand this idea, one should observe that
quasi-long-ranged CDW correlations in a single ladder
become long-ranged  in an array of coupled ladders 
because of the
pinning between the CDW's on different ladders. 
 Pinning  can occur
 only if the wavelength $\pi/(2 K_F)$ of the CDW 
is commensurate (equal) between 
 neighboring ladders. 

Thus,
CDW pinning can be frustrated 
by introducing
different Fermi momenta on the
ladders.
This is readily obtained, e. g.,
by introducing
 an alternating on-site energy offset $\eps$~\cite{ar.fr.03u}, 
as shown in Fig.~\ref{lad4}.
Interestingly, the model shown in Fig.~\ref{lad4}
displays the stripe structure which has been observed in several
high-Tc superconductors in various regions of their phase
diagram~\cite{experiments}. In particular, the period $4$ structure
seems to be the most stable stripe structure, at least in the
\LNSCO compound. 

The energy offset $\eps$ introduces
two 
different Fermi momenta $K_{F1}$ and $K_{F2}$, with
$K_{F1}- K_{F2} \sim \eps/ v_F$, which are associated to
 two
potentially relevant CDW's with different wavelengths.
The RG flux of the associated couplings $V_1$ and $V_2$ are both given
by the first of \eqref{rg} up to $\tau= \bar \tau \equiv \sim \ln \Delta_s /
\eps$.
For $\tau > \bar \tau$ the RG equation is given by
(we take for simplicity the same $K$ on the two different kinds of ladders)
\beq
\label{rgm}
{d \cal V_{1/2}}{d \tau} = (1 - K/2) {\cal V_{1/2}} \quad\quad \:.
\eeq
This form is due  to the fact that the $V_i$ operator ($i=1,2$) 
coupling CDW on neighboring ladders
continues to
increase only on the ladder with $K_F=K_{Fi}$, while it is frozen on
the other.

Combining \eqref{rg} with \eqref{rgm} it is easy to see that
the condition for $J$ to become more relevant  than
$V$, i. e. 
for superconductivity to win against CDW order
 is~\cite{ar.fr.03u} (we consider here the case of a not too small $\eps$) 
\beq
K > K_c = \sqrt{3} -1  \approx 0.8 \;,
\eeq
i. e. one has SC also for values of $K<1$.
Under this condition, the SC (Kosterlitz-Thouless~\cite{kt}) critical
temperature $T_c$ is obtained by integration of \eqref{rg} up to
$J(\tau^*)\sim 1$. One obtains~\cite{dop}
\beq
T_c \sim \Delta_s {\cal J_0}^{K/(2K-1)} \sim \Delta_s \left(\frac{
    \delt^2}{\Delta_s^2}\right)^{K/(2K-1)} \;.
\eeq
Notice that, in contrast to the BCS case in which $T_c$ is
exponentially small in the small coupling parameter
(in that case given by the effective attractive interaction), here $T_c$ is
only power-law small in $\delt$.

In conclusion, we have presented a model in which 
a relatively ``high'' $T_c$ is obtained by a purely repulsive
interaction between the electrons.
The model consists 
 of an array of weakly-coupled two-leg ladders with alternating
 on-site energies. This alternation  is important in order to 
prevent the system to develop into a CDW insulator before 
becoming a superconductor.

This work was supported, in part, 
by NSF grants No. DMR 01-10329 
at UCLA (SAK), and DMR-01-32990 at the University of Illinois (EF), 
and by a Heisenberg grant (AR 324/3-1) form the DFG (EA).




\begin{thebibliography}{10}
\bibitem{dop}
This holds away from special values of the doping, where $\Delta_s$ is of the
  order of the exchange coupling. Special dopings are discussed in
  Ref.~\cite{ar.fr.03u}.
\bibitem{kt}
In two dimensions the order is long ranged only at $T=0$, otherwise one obtains
  a quasi-long-ranged phase below the Kosterlitz-Thouless temperature.
\end{thebibliography}


%
%
%
%


\end{document}